\documentclass[conference]{IEEEtran}
\IEEEoverridecommandlockouts
\usepackage{graphicx,cite,epsfig,amssymb,amsmath,color}
\usepackage{subfigure}
\usepackage{float}
\usepackage{algorithm}
\usepackage{algpseudocode}
\usepackage{cuted}
\usepackage{stfloats}
\usepackage{comment}
\usepackage{ragged2e}

\def\BibTeX{{\rm B\kern-.05em{\sc i\kern-.025em b}\kern-.08em
    T\kern-.1667em\lower.7ex\hbox{E}\kern-.125emX}}
\begin{document}

\title{Multi-User Delay Alignment Modulation for Millimeter Wave Massive MIMO}

\author{
	\IEEEauthorblockN{
		Xingwei Wang\IEEEauthorrefmark{1}, 
		Haiquan Lu\IEEEauthorrefmark{1}\IEEEauthorrefmark{2},
		Yong Zeng\IEEEauthorrefmark{1}\IEEEauthorrefmark{2}} 
	\IEEEauthorblockA{\IEEEauthorrefmark{1}National Mobile Communications Research Laboratory, Southeast University, Nanjing 210096, China}
	\IEEEauthorblockA{\IEEEauthorrefmark{2}Purple Mountain Laboratories, Nanjing 211111, China\\Email: \{xingwei-wang, haiquanlu, yong\_zeng\}@seu.edu.cn}
}

\maketitle

\begin{abstract}
Delay alignment modulation (DAM) is a novel wideband communication technique, which exploits the high spatial resolution and multi-path sparsity of millimeter wave (mmWave) massive multiple-input multiple-output (MIMO) systems to mitigate inter-symbol interference (ISI), without relying on conventional techniques like channel equalization or multi-carrier transmission. In this paper, we extend the DAM technique to multi-user mmWave massive MIMO communication systems. We first provide asymptotic analysis by showing that when the number of base station (BS) antennas is much larger than the total number of channel paths, DAM is able to eliminate both ISI and inter-user interference (IUI) with the simple delay pre-compensation and per-path-based maximal ratio transmission (MRT) beamforming. We then study the general multi-user DAM design by considering the three classical transmit beamforming strategies in a per-path basis, namely MRT, zero-forcing (ZF) and regularized zero-forcing (RZF). Simulation results demonstrate that  multi-user DAM can significantly outperform the benchmarking single-carrier ISI mitigation technique that only uses the strongest channel path of each user.
\end{abstract}

\IEEEpeerreviewmaketitle

\addtolength{\topmargin}{0.02in}

\section{Introduction}

Recently, delay alignment modulation (DAM) was proposed as a novel technique to tackle the inter-symbol interference (ISI) issue, without relying on conventional techniques like channel equalization or multi-carrier transmission \cite{1}. Specifically, by  leveraging the super spatial resolution of large antenna arrays \cite{2} and the inherent multi-path sparsity of high frequency channels like millimeter wave (mmWave)\cite{3} and Terahertz channels, DAM enables manipulable channel delay spread by means of {\it delay compensation} and {\it path-based beamforming} \cite{1}. As a result, all the multi-path signal components may reach the receiver concurrently and constructively, rather than causing the detrimental ISI. This renders DAM resilient to the time-dispersive channel for more efficient single- or multi-carrier transmissions. Some preliminary works on DAM are presented in \cite{4,5,6,7,8}. For example, in \cite{4}, an efficient channel estimation method was developed for DAM communication. By combining DAM with orthogonal frequency division multiplexing (OFDM), a novel DAM-OFDM scheme was proposed in \cite{5}, which may outperform OFDM in terms of spectral efficiency, bit error rate (BER), and peak-to-average-power ratio (PAPR). The investigations  of DAM for  integrated sensing and communication (ISAC) and multiple-intelligent reflecting surfaces (IRSs) aided communication were investigated in \cite{6}, \cite{7} and \cite{8}, respectively.

DAM is significantly different from the existing ISI-mitigation techniques developed over the fast few decades,  such as channel equalization and  multi-carrier OFDM transmission. For example, typical time-domain equalization techniques include time reversal (TR)\cite{9,10,11} and channel shortening\cite{12,13}. Specifically, TR mainly treats the multi-path channel as an intrinsic matched filter\cite{9,10,11} and addresses the ISI issue via the rate back-off technique\cite{9,10}. In \cite{12} and \cite{13}, channel shortening technique was proposed by applying a  short time-domain finite impulse response (FIR) filter to shorten the effective channel impulse response in order to avoid the use of long cyclic prefix (CP). Besides, OFDM is the dominating wideband communication technology, though it suffers from well-known practical issues like carrier frequency offset (CFO), high PAPR, and the severe out-of-band (OOB) emission\cite{14}. By contrast, DAM is a spatial-delay processing technique, which utilizes  the high spatial  resolution  and  multi-path sparsity of mmWave 
massive multiple-input multiple-output (MIMO) systems, while circumventing the aforementioned issues suffered by conventional channel equalization or multi-carrier communication. It is worth noting that relevant  techniques for delay spread reduction were studied in \cite{15} and \cite{16}. However, neither of them  exploits the  high spatial resolution or multi-path sparsity of mmWave massive MIMO systems to completely eliminate ISI. In addition, both our prior work \cite{17} and \cite{18} apply delay compensation in the mmWave MIMO systems, but they were designed only for the special lens MIMO or hybrid beamforming architectures.

Note that the existing works\cite{1,4,5,6,7,8} on DAM only focus on the single user scenario. In this paper, we extend the DAM technique to the multi-user  mmWave massive MIMO systems. To gain the essential insights, we first provide an asymptotic analysis by assuming that the number of BS antennas $M_t$ is much larger than the total number of  channel multi-paths $L_{\mathrm{tot}}$. In this case, we show that the time-dispersive multi-user channel can be transformed into ISI-free and inter-user interference (IUI)-free additive white Gaussian noise (AWGN) channels with the simple delay pre-compensation and per-path-based MRT beamforming. Then for the general scenario with given finite number of BS antennas, the  multi-user DAM design is investigated by considering three classical transmit beamforming strategies in a per-path basis, i.e., MRT, zero-forcing (ZF) and regularized zero-forcing (RZF). To be specific, it is shown than when $M_t\geq L_{\mathrm{tot}}$, both the ISI and IUI can be completely eliminated with the path-based ZF beamforming. Furthermore, the low complexity path-based MRT beamforming and the more general path-based RZF beamforming are  respectively developed for multi-user DAM with tolerable  residual ISI and IUI. Finally, simulation results are provided to  demonstrate the superiority of DAM to the benchmarking single-carrier ISI mitigation technique that only uses the strongest path of each user.

\begin{figure}
    \centering
    \includegraphics[scale=0.3]{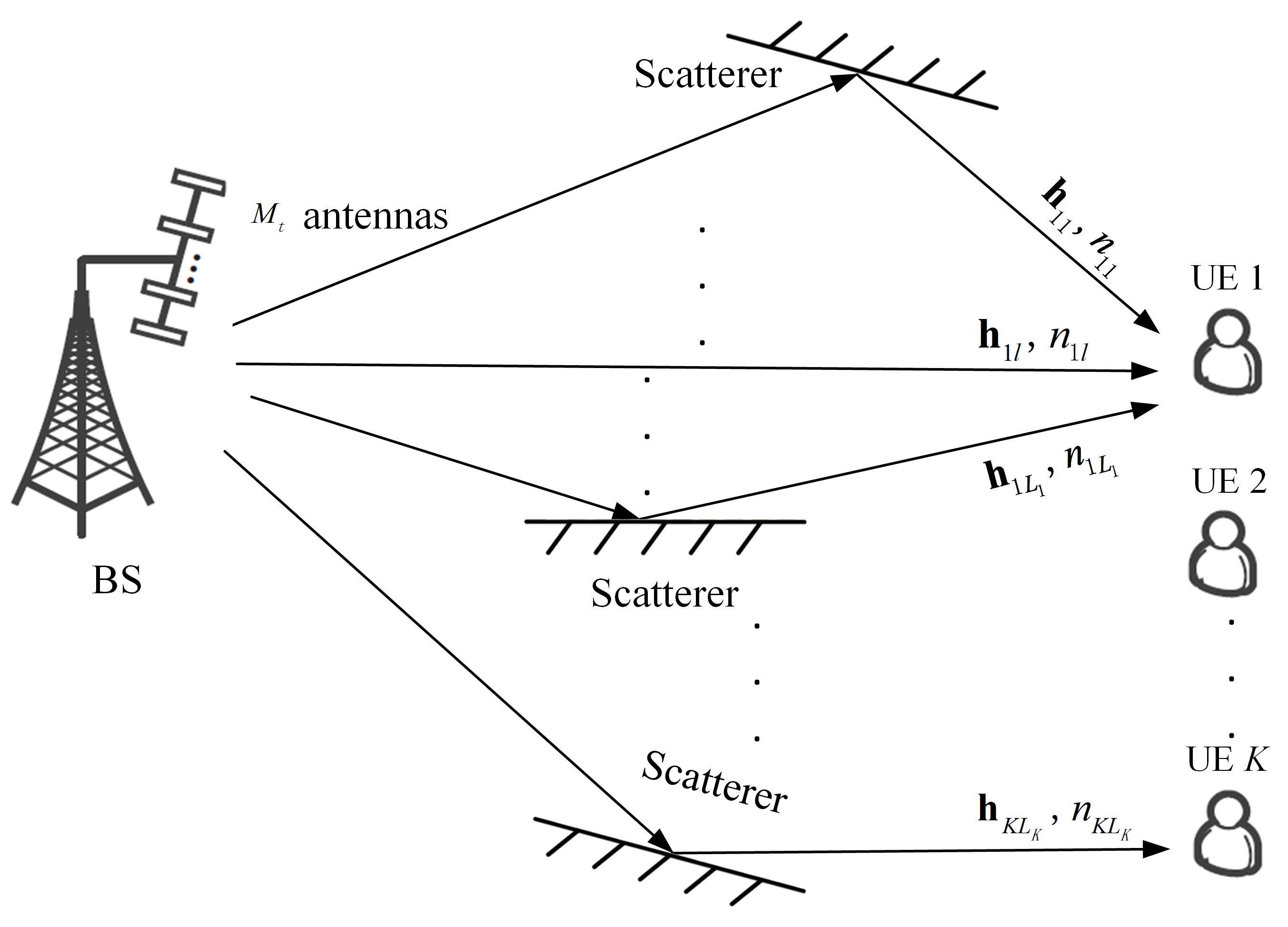}
    \caption{A multi-user downlink mmWave massive MIMO communication system in time-dispersive channels.} \label{fig1}
     \vspace{-3ex}
\end{figure}

\section{System Model and Asymptotic Analysis} \label{sec1}

As shown in Fig. \ref{fig1}, we consider a multi-user mmWave massive MIMO downlink communication system, where the BS equipped with $M_t \gg 1$ antennas serves $K$ single-antenna user equipments (UEs). In multi-path environment, the baseband discrete-time channel impulse response of UE $k$ for each channel coherence block can be expressed as
\setlength\abovedisplayskip{2.5pt}
\setlength\belowdisplayskip{2.5pt}
\begin{equation}
\boldsymbol{\bf{h}}_k[n]=\sum_{l=1}^{L_k} \boldsymbol{\bf{h}}_{kl} \delta[n-n_{kl}], \label{1}
\end{equation}
where $\boldsymbol{\bf{h}}_{kl} \in \mathbb{C}^{M_t \times 1}$ denotes the channel coefficient vector for the $l$th multi-path of UE $k$, $L_k$ is the number of temporal-resolvable multi-paths of UE $k$, and $n_{kl}$ denotes its discretized  delay. Let $n_{k, \text{max}}=\max\limits_{1\leq l\leq L_k}n_{kl}$ and $n_{k, \text{min}}=\min\limits_{1\leq l\leq L_k}n_{kl}$  denote the maximum  and minimum delays of UE $k$, respectively. 

\begin{figure}
\centering
	\includegraphics[scale=.39]{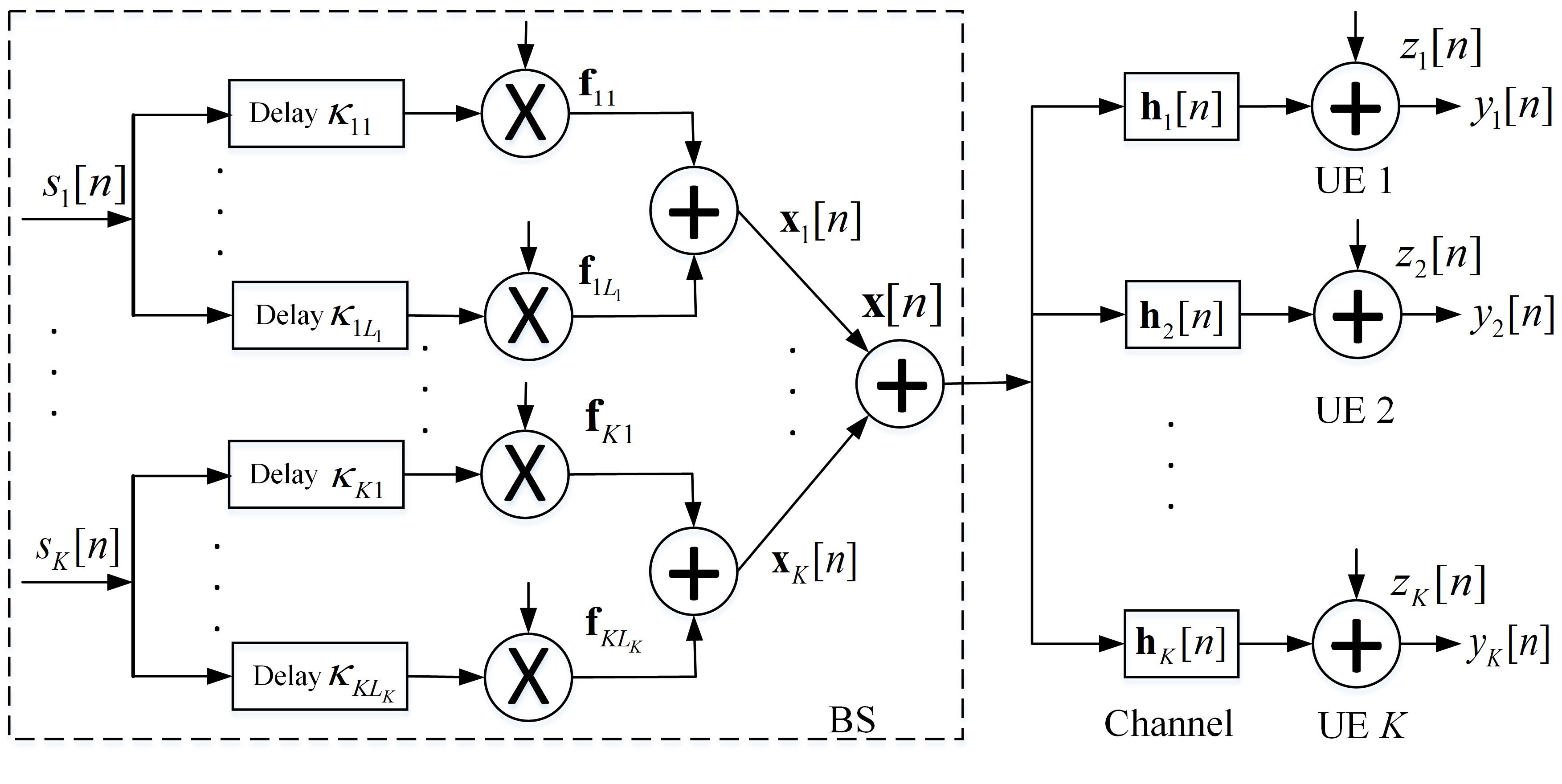}
\caption{Block diagram  for multi-user communication with delay alignment modulation.}\label{fig2}
 \vspace{-3ex}
\end{figure}

Let $s_k[n]$ be the independent and identically distributed (i.i.d.) information-bearing symbols of UE $k$ with normalized power $\mathbb{E}[|s_k[n]|^2]=1$. By extending the DAM technique proposed in \cite{1} to the multi-user scenario, the transmitted signal by the BS for multi-user DAM is
\begin{equation}
\boldsymbol{\bf{x}}[n]=\sum_{k=1}^{K}\sum_{l'=1}^{L_k}\boldsymbol{\bf{f}}_{kl'} s_{k}[n-\kappa_{kl'}], \label{211}
\end{equation}
where $\boldsymbol{\bf{f}}_{kl'} \in \mathbb{C}^{M_t \times 1}$ denotes the per-path based transmit beamforming vector associated with multi-path $l'$ of UE $k$, and $\kappa_{kl'}$  is the deliberately introduced delay pre-compensation, which is set as $\kappa_{kl'}=n_{k, \text{max}}-n_{kl'} \geq 0, \forall l'=1,...,L_{k}$. The block diagram  for  multi-user DAM is given in Fig. \ref{fig2}. By using the fact that $s_{k}[n]$ is independent across different $n$ and $k$ as well as  $\kappa_{kl'}\neq \kappa_{kl}, \forall l\neq l'$, it can be shown that the transmit power of the BS is
\begin{equation}
\mathbb{E}[\|\boldsymbol{\bf{x}}[n]\|^2]=\sum_{k=1}^K \sum_{l'=1}^{L_{k}}  \|\boldsymbol{\bf{f}}_{kl'}\|^2 \leq P, \label{333}
\end{equation}
where $P$ is the maximum allowable transmit power.

Based on \eqref{1} and \eqref{211}, the received signal of UE $k$ is
\begin{align}
&y_k[n]=\boldsymbol{\bf{h}}_k^H[n]*\boldsymbol{\bf{x}}[n]+ z_k[n]\label{55} \\  \notag
&=\underbrace{\left(\sum_{l=1}^{L_k}\boldsymbol{\bf{h}}_{kl}^H\boldsymbol{\bf{f}}_{kl}\right)  s_k[n-n_{k, \text{max}}]}_{\text{Desired signal}} \\  \notag
                 & +\underbrace{\sum_{l=1}^{L_k} \sum_{l'\ne l}^{L_k} \boldsymbol{\bf{h}}_{kl}^H \boldsymbol{\bf{f}}_{kl'} s_k[n-n_{k, \text{max}}-n_{kl}+n_{kl'}] }_{\text{ISI}} \\  \notag
                  & +\underbrace{\sum_{l=1}^{L_k}\sum_{k'\ne k}^{K}\sum_{l'=1}^{L_{k'}}\boldsymbol{\bf{h}}_{kl}^H  \boldsymbol{\bf{f}}_{k'l'}s_{k'}[n-n_{k', \text{max}}-n_{kl}+n_{k'l'}]}_{\text{IUI}} +z_k[n],
\end{align}
where $*$ denotes linear convolution, and $z_k[n]\sim \mathcal{CN}(0,\sigma^2)$ is the AWGN. It is observed that if UE $k$ is locked to the delay $n_{k, \text{max}}$, then the first term is the desired signal, and the second and third terms are the ISI and IUI, respectively.  

Fortunately, we show that in mmWave massive MIMO systems when the number of BS antennas $M_t$ is much larger than the total number of multi-paths $L_{\mathrm{tot}}=\sum_{k=1}^K L_{k}$, both the ISI and IUI in \eqref{55} asymptotically vanish  with the simple path-based MRT beamforming. To this end,  we first study the correlation property of the channel vectors $\boldsymbol{\bf{h}}_{kl}, k=1,...,K, l=1,...,L_k$. For ease of exposition, the basic uniform linear array (ULA) with adjacent elements separated by half wavelength is considered, for which $\boldsymbol{\bf{h}}_{kl}$ in \eqref{1} can be written  as
\begin{equation}
\boldsymbol{\bf{h}}_{kl}= \alpha_{kl}\boldsymbol{\bf{a}}(\theta_{kl})=\alpha_{kl}[1,e^{-j\pi \cos\theta_{kl}},...,e^{-j\pi(M_t-1)\cos\theta_{kl}}]^T, \label{611}\\
\end{equation}
where $\alpha_{kl}$ denotes the complex-valued path gain for the $l$th multi-path of UE $k$, and $\theta_{kl}$ and $\boldsymbol{\bf{a}}(\theta_{kl})\in \mathbb{C}^{M_t\times1}$ denote the angle of departure (AoD) and the transmit array response vector of multi-path $l$ of UE $k$, respectively. Note that for ease of the asymptotical analysis, we assume that each temporal-resolvable multi-path corresponds to one AoD. It follows from \cite{5} that as long as the multi-paths correspond to distinct AoDs, i.e., $\theta_{kl}\neq \theta_{k'l'}, \forall k\neq k'$ or $l\neq l'$, the transmit array response vectors are asymptotically orthogonal when $M_t \gg L_{\mathrm{tot}}$, i.e., 
\begin{equation}
\lim_{M_t \to \infty}\frac{|\boldsymbol{\bf{h}}_{kl}^H\boldsymbol{\bf{h}}_{k'l'}|}{\|\boldsymbol{\bf{h}}_{kl}\|~\|\boldsymbol{\bf{h}}_{k'l'}\|} \to 0, \forall k'\ne k, \text{or}~ l'\ne l.\label{kk}
\end{equation}

Consider the low-complexity path-based MRT beamforming  $ \boldsymbol{\bf{f}}_{kl'}=\xi_k\sqrt{p_k}\boldsymbol{\bf{h}}_{kl'}$, where $\xi_k=1/\sqrt{\sum_{j=1}^{L_k}\|\boldsymbol{\bf{h}}_{kj}\|^2}$ is the normalization factor and $p_k$ denotes the power allocated to UE $k$. In this case, by scaling the signal in \eqref{55} with $\xi_k$, we have 
\begin{align}
\xi_k y_k[n]&=\sqrt{p_k} s_k[n-n_{k, \text{max}}]  +\sqrt{p_k}\sum_{l=1}^{L_k} \sum_{l'\ne l}^{L_k} \frac{\boldsymbol{\bf{h}}_{kl}^H \boldsymbol{\bf{h}}_{kl'}}{\sum_{j=1}^{L_k}\|\boldsymbol{\bf{h}}_{kj}\|^2}\label{5mm5} \nonumber \\  
                  &\times s_k[n-n_{k, \text{max}}-n_{kl}+n_{kl'}] \nonumber \\ 
                  & +\sum_{l=1}^{L_k}\sum_{k'\ne k}^{K}\sum_{l'=1}^{L_{k'}} \frac{\sqrt{p_{k'}}\boldsymbol{\bf{h}}_{kl}^H \boldsymbol{\bf{h}}_{k'l'}}{\sqrt{\sum_{j=1}^{L_k}\|\boldsymbol{\bf{h}}_{kj}\|^2}~ \sqrt{\sum_{j=1}^{L_{k'}}\|\boldsymbol{\bf{h}}_{k'j}\|^2}}
              \nonumber \\ 
                  &\times s_{k'}[n-n_{k', \text{max}}-n_{kl}+n_{k'l'}] +\xi_k z_k[n].
\end{align}
Thanks to the asymptotically orthogonal property in \eqref{kk}, when $M_t \gg L_{\mathrm{tot}}$, we have 
\begin{equation}
\frac{|\boldsymbol{\bf{h}}_{kl}^H \boldsymbol{\bf{h}}_{kl'}|}{\sum_{j=1}^{L_k}\|\boldsymbol{\bf{h}}_{kj}\|^2} \leq \frac{|\boldsymbol{\bf{h}}_{kl}^H \boldsymbol{\bf{h}}_{kl'}|}{\|\boldsymbol{\bf{h}}_{kl}\|^2+\|\boldsymbol{\bf{h}}_{kl'}\|^2}\leq \frac{|\boldsymbol{\bf{h}}_{kl}^H \boldsymbol{\bf{h}}_{kl'}|}{\|\boldsymbol{\bf{h}}_{kl}\|~\|\boldsymbol{\bf{h}}_{kl'}\|}\to 0, \label{my11}
\end{equation}
and
\begin{equation}
\begin{split}
&\frac{|\boldsymbol{\bf{h}}_{kl}^H \boldsymbol{\bf{h}}_{k'l'}|}{\sqrt{(\sum_{j=1}^{L_k}\|\boldsymbol{\bf{h}}_{kj}\|^2) (\sum_{j=1}^{L_{k'}}\|\boldsymbol{\bf{h}}_{k'j}\|^2)}} \label{my22} \\  
&\leq \frac{|\boldsymbol{\bf{h}}_{kl}^H \boldsymbol{\bf{h}}_{k'l'}|}{\sqrt{\|\boldsymbol{\bf{h}}_{kl}\|^2~ \|\boldsymbol{\bf{h}}_{k'l'}\|^2}}= \frac{|\boldsymbol{\bf{h}}_{kl}^H \boldsymbol{\bf{h}}_{k'l'}|}{\|\boldsymbol{\bf{h}}_{kl}\|~ \|\boldsymbol{\bf{h}}_{k'l'}\|} \to 0. 
\end{split}
\end{equation}
With \eqref{my11} and \eqref{my22}, and by dividing both sides of \eqref{5mm5} with $\xi_k$, the resulting signal for UE $k$ reduces to
\begin{equation}
y_k[n]\to \sqrt{p_k\begin{matrix}\sum_{j=1}^{L_k}\end{matrix}\|\boldsymbol{\bf{h}}_{kj}\|^2} s_k[n-n_{k, \text{max}}] +z_k[n].  \label{dam1}            
\end{equation}

It is observed from \eqref{dam1} that the resulting signal of UE $k$ only includes the symbol sequence $s_k[n]$ with one single delay $n_{k,\text{max}}$, while achieving the multiplicative gain contributed  by all the  $L_k$ multi-paths. As a result, the original time-dispersive multi-user interfering  channel  has been transformed to $K$ parallel  ISI- and IUI-free AWGN channels, without relying on  the conventional techniques like  channel equalization or multi-carrier transmission. Moreover, the signal-to-noise
ratio (SNR) in \eqref{dam1} is given by $\gamma_k^{\text{MRT}} ={p_k\|\bar{\boldsymbol{\bf{h}}}_{k}\|^2}/{\sigma^2}$, where $\bar{\boldsymbol{\bf{h}}}_{k} = [\boldsymbol{\bf{h}}_{k1}^H,...,\boldsymbol{\bf{h}}_{kL_k}^H]^H \in \mathbb{C}^{M_tL_k\times1}$. The optimal power allocation to maximize the asymptotic sum rate can be obtained by the classical water-filling (WF) power allocation.

\section{Path-Based Beamforming  for Multi-User DAM}
In this section, we consider the practical scenario with given finite number of antennas $M_t$, where three classical  precoding schemes, namely MRT, ZF and RZF are studied  on the per-path basis to cater for multi-user DAM transmission.

To derive the signal-to-interference-plus-noise ratio (SINR) for the general signal in \eqref{55}, the symbols need to be properly grouped by considering the delay differences of the signal components\cite{1}. To this end, let $\Delta_{kl,k'l'}=n_{kl}-n_{k'l'}$  denote the {\it delay difference} between multi-path $l$ of UE $k$ and multi-path $l'$ of UE $k'$. Then for a given UE pair $(k,k')$, we have $ \Delta_{kl,k'l'}\in \{\Delta_{kk',\min},\Delta_{kk',\min}+1,...,\Delta_{kk',\max} \}$, where $\Delta_{kk',\max} =n_{k,\text{max}}- n_{k', \text{min}}$ and $\Delta_{kk',\min} =n_{k, \text{min}}- n_{k', \text{max}}$. Then for each UE pair $(k,k')$ and delay difference $i$ with $i\in  \{\Delta_{kk',\min},\Delta_{kk',\min}+1,...,\Delta_{kk',\max} \}$, we have the following effective channel
 \begin{equation}
\boldsymbol{\bf{g}}_{kk'l'} [i] = 
\begin{cases}
\boldsymbol{\bf{h}}_{kl}, &\text{if}~\exists l\in \{1,...,L_k\}, ~\mbox{s.t.}~\Delta_{kl,k^{'}l^{'}}=i, \\ 
\boldsymbol{\bf{0}} , &\mbox{otherwise}.
\end{cases}
\end{equation}

Therefore, \eqref{55} can be equivalently written as 
\begin{align}
&y_k[n]={\left(\sum_{l=1}^{L_k}\boldsymbol{\bf{h}}_{kl}^H \boldsymbol{\bf{f}}_{kl}\right) s_k[n-n_{k, \text{max}}]}  \label{2l5} \\  \notag
                 & +{ \sum_{i=\Delta_{kk,\min},i\ne 0}^{\Delta_{kk,\max}}  \left(\sum_{l'=1}^{L_k}\boldsymbol{\bf{g}}_{kkl'}^H[i] \boldsymbol{\bf{f}}_{kl'}\right) s_k[n-n_{k, \text{max}}-i]} \\  \notag
                 &+{\sum_{k'\ne k}^{K}\sum_{i=\Delta_{kk',\min}}^{\Delta_{kk',\max}}\left(\sum_{l'=1}^{L_{k'}} \boldsymbol{\bf{g}}_{kk'l'}^H[i] \boldsymbol{\bf{f}}_{k'l'}\right)  s_{k'}[n-n_{k', \text{max}}-i] }\\  \notag
                 & +z_k[n].
\end{align}
The resulting SINR is given  in~\eqref{45}, shown at the top of the next page, where we have defined $\bar{\boldsymbol{\bf{f}}}_{k}=[\boldsymbol{\bf{f}}_{k1}^H,...,\boldsymbol{\bf{f}}_{kL_{k}}^H]^H\in \mathbb{C}^{M_tL_{k}\times1}, \bar{\boldsymbol{\bf{g}}}_{kk'}[i]=[\boldsymbol{\bf{g}}_{kk'1}^H[i],...,\boldsymbol{\bf{g}}_{kk'L_{k'}}^H[i]]^H\in \mathbb{C}^{M_tL_{k'} \times 1 }$,  $\boldsymbol{\bf{G}}_{kk'} = [\bar{\boldsymbol{\bf{g}}}_{kk'}[\Delta_{kk',\min}],...,\bar{\boldsymbol{\bf{g}}}_{kk'}[\Delta_{kk',\max}]] \in \mathbb{C}^{M_tL_{k'} \times {\Delta_{kk', \text{span}}} }$, with  $ \Delta_{kk', \text{span}}=\Delta_{kk',\max} -\Delta_{kk',\min}$.
\newcounter{TempEqCnt} 
\setcounter{TempEqCnt}{\value{equation}} 
\setcounter{equation}{12} 
\begin{figure*}[ht] 
\begin{equation}
\begin{split}
\gamma_k&= \frac{|\begin{matrix}\sum_{l=1}^{L_k}\end{matrix}\boldsymbol{\bf{h}}_{kl}^H \boldsymbol{\bf{f}}_{kl}|^2}{ \sum_{i=\Delta_{kk,\min},i\ne 0}^{\Delta_{kk,\max}}  |\sum_{l'= 1}^{L_k} \boldsymbol{\bf{g}}_{kkl'}^H[i]\boldsymbol{\bf{f}}_{kl'} |^2+ \sum_{k'\ne k}^K\sum_{i=\Delta_{kk',\min}}^{\Delta_{kk',\max}}   |\sum_{l'=1}^{L_{k'}}\boldsymbol{\bf{g}}_{kk'l'}^H[i] \boldsymbol{\bf{f}}_{k'l'}|^2+\sigma^2 } \label{45}\\ 
&=\frac{|\bar{\boldsymbol{\bf{h}}}_k^H\bar{\boldsymbol{\bf{f}}}_k|^2}{\sum_{i=\Delta_{kk,\min},i\ne 0}^{\Delta_{kk,\max}} |\bar{\boldsymbol{\bf{g}}}_{kk}^H[i]\bar{\boldsymbol{\bf{f}}}_k|^2 +\sum_{k'\ne k}^{K} \sum_{i=\Delta_{kk',\min}}^{\Delta_{kk',\max}}  |\bar{\boldsymbol{\bf{g}}}_{kk'}^H[i]\bar{\boldsymbol{\bf{f}}}_{k'}|^2 + \sigma^2 }\\  
&=\frac{|\bar{\boldsymbol{\bf{h}}}_k^H\bar{\boldsymbol{\bf{f}}}_k|^2}{ ||\boldsymbol{\bf{G}}_{kk}^H\bar{\boldsymbol{\bf{f}}}_k||^2 +\sum_{k'\ne k}^{K} ||\boldsymbol{\bf{G}}_{kk'}^H\bar{\boldsymbol{\bf{f}}}_{k'}||^2+\sigma^2 }.
\end{split}
\end{equation}
\hrulefill
\end{figure*}

\vspace{-0.4cm}
\subsection{Path-Based MRT Beamforming}
The path-based MRT beamforming for asymptotic analysis with $M_t\gg L_{\mathrm{tot}}$ has been given in Section \ref{sec1}. For the practical setup with finite $M_t$, the low complexity path-based MRT beamforming for multi-path $l$ of UE $k$ is given by
\begin{equation}
\boldsymbol{\bf{f}}_{kl}^{\text{MRT}}=\sqrt{P}{\boldsymbol{\bf{h}}_{kl}}\big/{\|\boldsymbol{\bf{H}}\|_{F}},\label{ll1}
\end{equation}
where $\boldsymbol{\bf{H}}=[\boldsymbol{\bf{h}}_{11,}...,\boldsymbol{\bf{h}}_{1L_1},...,\boldsymbol{\bf{h}}_{K1,}...,\boldsymbol{\bf{h}}_{KL_K}] \in \mathbb{C}^{M_t \times L_{\mathrm{tot}}}$. Denote by $\bar{\boldsymbol{\bf{f}}}_{k}^{\text{MRT}}=\sqrt{P}{\bar{\boldsymbol{\bf{h}}}_k}/{\|\boldsymbol{\bf{H}}\|_{F}}$, with $\bar{\boldsymbol{\bf{f}}}_{k}^{\text{MRT}}$ $=[(\boldsymbol{\bf{f}}_{k1}^{\text{MRT}})^H,...,(\boldsymbol{\bf{f}}_{kL_k}^{\text{MRT}})^H]^H$, and by substituting it into \eqref{45}, the resulting SINR can be expressed as
\begin{equation}
\gamma_{k}^{\text{MRT}}=\frac{\|\bar{\boldsymbol{\bf{h}}}_k\|^4}{ ||\boldsymbol{\bf{G}}_{kk}^H\bar{\boldsymbol{\bf{h}}}_k||^2 +\sum_{k'\ne k}^{K} ||\boldsymbol{\bf{G}}_{kk'}^H\bar{\boldsymbol{\bf{h}}}_{k'}||^2+\|\boldsymbol{\bf{H}}\|_{F}^2\sigma^2/P }.
\end{equation}

\subsection{Path-Based ZF Beamforming} \label{sec2}
With path-based ZF beamforming, $\boldsymbol{\bf{f}}_{kl'}, k = 1,...,K, l'=1,...,L_k,  $ are designed so that the ISI and IUI in \eqref{55} are all eliminated, i.e., 
\begin{equation}
\boldsymbol{\bf{h}}_{kl}^H \boldsymbol{\bf{f}}_{kl'}^{\text{ZF}}=0, ~\forall l'\neq l, \\ \label{ll2}
\end{equation}
\begin{equation}
\boldsymbol{\bf{h}}_{kl}^H \boldsymbol{\bf{f}}_{k'l'}^{\text{ZF}}=0, ~\forall k'\neq k, \text{and} ~\forall l, l'. \label{ll3}
\end{equation}
Denote by $\boldsymbol{\bf{H}}_{kl'}\in \mathbb{C}^{M_t \times (L_{\mathrm{tot}}-1)}$  the submatrix of $\boldsymbol{\bf{H}}$ by  excluding the column $\boldsymbol{\bf{h}}_{kl'}$. Thus, the path-based ZF constraints in \eqref{ll2} and \eqref{ll3} can be compactly written as
\begin{equation}
\boldsymbol{\bf{H}}_{kl'}^H \boldsymbol{\bf{f}}_{kl'}^{\text{ZF}}=\boldsymbol{\bf{0}}_{(L_{\mathrm{tot}}-1)\times 1}, ~\forall k, l'. \\ \label{aa2}
\end{equation}
The above ZF constraint is feasible when $M_t\geq L_{\mathrm{tot}}$.

Let $\boldsymbol{\bf{F}}^{\text{ZF}} =[\boldsymbol{\bf{f}}_{11}^{\text{ZF}},...,\boldsymbol{\bf{f}}_{1L_1}^{\text{ZF}},...,\boldsymbol{\bf{f}}_{K1}^{\text{ZF}},...,\boldsymbol{\bf{f}}_{KL_K}^{\text{ZF}}]\in \mathbb{C}^{M_t \times L_{\mathrm{tot}}}$ denote the matrix composed by all the path-based ZF beamforming vectors. Without loss of generality, we may decompose  $\boldsymbol{\bf{F}}^{\text{ZF}}$ as $\boldsymbol{\bf{F}}^{\text{ZF}}=\boldsymbol{\bf{W}}\boldsymbol{\bf{V}}^{\frac{1}{2}}$, where $\boldsymbol{\bf{W}} =[{\boldsymbol{\bf{w}}}_{11},...,{\boldsymbol{\bf{w}}}_{1L_1},...,{\boldsymbol{\bf{w}}}_{K1},...,{\boldsymbol{\bf{w}}}_{KL_K}]\in \mathbb{C}^{M_t \times L_{\mathrm{tot}}}$ is designed to guarantee the ZF constraints in \eqref{aa2}, and $\boldsymbol{\bf{V}}=\text{diag}\{v_{11},...,v_{1L_1},...,v_{K1},...,v_{KL_K}\}$ containing non-negative real-valued diagonal elements is  the power allocation matrix ensuring the power constraint $\sum_{k=1}^K \sum_{l'=1}^{L_{k}}  \|\boldsymbol{\bf{f}}_{kl'}^{\text{ZF}}\|^2 = P $. One effective solution for $\boldsymbol{\bf{W}}$ to guarantee the ZF constraints  \eqref{aa2} is by letting $\boldsymbol{\bf{H}}^H\boldsymbol{\bf{W}}=\boldsymbol{\bf{I}}_{L_{\mathrm{tot}}}$. When $M_t\geq L_{\mathrm{tot}}$, $\boldsymbol{\bf{W}}$ can be directly obtained by taking the right pseudo inverse of $\boldsymbol{\bf{H}}^H$, i.e., $\boldsymbol{\bf{W}}=(\boldsymbol{\bf{H}}^H)^{\dagger} = \boldsymbol{\bf{H}}(\boldsymbol{\bf{H}}^H\boldsymbol{\bf{H}})^{-1}$. As a result, the path-based ZF beamforming $\boldsymbol{\bf{f}}_{kl}^{\text{ZF}}$ can be expressed as $\boldsymbol{\bf{f}}_{kl}^{\text{ZF}} =\sqrt{v_{kl}}\boldsymbol{\bf{w}}_{kl}$, where $\boldsymbol{\bf{w}}_{kl}$ is the $\left((\sum_{j=1}^{k-1}L_j)+l\right)$th column of $\boldsymbol{\bf{W}}$. By substituting $\boldsymbol{\bf{f}}_{kl}^{\text{ZF}}$ into \eqref{55}, the received signal reduces to 
\begin{align} 
y_k[n]&=\begin{matrix}\sum_{l=1}^{L_k}\end{matrix}\sqrt{v_{kl}} \boldsymbol{\bf{h}}_{kl}^H\boldsymbol{\bf{w}}_{kl} s_k[n-n_{k, \text{max}}] + z_k[n] \label{oo}  \\ \notag
&=\left(\begin{matrix}\sum_{l=1}^{L_k}\end{matrix}\sqrt{v_{kl}}\right) s_k[n-n_{k, \text{max}}] + z_k[n],
\end{align}
where we have used the identity  $\boldsymbol{\bf{h}}_{kl}^H\boldsymbol{\bf{w}}_{kl}=1$ based on the property of pseudo inverse. It is observed from \eqref{oo} that similar to \eqref{dam1}, the original time-dispersive multi-user interfering channel is also transformed into $K$ parallel  ISI- and IUI-free AWGN channels. The SNR of UE $k$  is $\gamma_{k}^{\text{ZF}}={(\sum_{l=1}^{L_k}\sqrt{v_{kl}})^2 }\big/{ \sigma^2}$, 
and the achievable rate of UE $k$ in bits/second/Hz (bps/Hz) is given by $\text{log}_2(1+\gamma_{k}^{\text{ZF}})$.
The optimal power allocation coefficients $v_{kl}, \forall k,l$, to  maximize the sum rate can be found by solving the following problem
\begin{equation}
\begin{split}
\max\limits_{{v}_{kl}, \forall k,l}~ &\begin{matrix}\sum_{k=1}^K\end{matrix} \text{log}_2\left(1+{(\begin{matrix}\sum_{l=1}^{L_k}\end{matrix}\sqrt{v_{kl}})^2 }\big/{ \sigma^2}\right)\label{o22o}\\
\text{s.t.}~&\begin{matrix}\sum_{k=1}^K \end{matrix} \begin{matrix}\sum_{l=1}^{L_k}\end{matrix}v_{kl}\|\boldsymbol{\bf{w}}_{kl}\|^2 \leq P,\\
& {v}_{kl}\geq 0, \forall k,l.
\end{split}
\end{equation}
By defining $\bar{v}_{kl}={v_{kl}}\|\boldsymbol{\bf{w}}_{kl}\|^2$, $\boldsymbol{\bf{t}}_{k} =[\sqrt{\bar{v}_{k1}},...,\sqrt{\bar{v}_{kL_k}}]^T$, and $\boldsymbol{\bf{q}}_{k} =[{1}/{\|\boldsymbol{\bf{w}}_{k1}\|},...,{1}/{\|\boldsymbol{\bf{w}}_{kL_k}\|}]^T$, \eqref{o22o} can be equivalently written as
\begin{equation}
\begin{split}
\max\limits_{\{\boldsymbol{\bf{t}}_k\}_{k=1}^K}~ &\begin{matrix}\sum_{k=1}^K\end{matrix} \text{log}_2\left(1+{(\boldsymbol{\bf{t}}_{k}^T\boldsymbol{\bf{q}}_{k})^2 }\big/{ \sigma^2}\right)\label{o2o}\\
\text{s.t.}~&\begin{matrix}\sum_{k=1}^K\end{matrix}  \|{\boldsymbol{\bf{t}}_{k}}\|^2 \leq P, \\
& {\boldsymbol{\bf{t}}_{k}}\succeq  \boldsymbol{0}, \forall k.
\end{split}
\end{equation}

To derive the optimal solution to \eqref{o2o}, the auxiliary variables $\{P_k\}_{k=1}^K$ are introduced, so that \eqref{o2o} can be equivalently written as 
\begin{equation}
\begin{split}
\max\limits_{\{\boldsymbol{\bf{t}}_k, P_k\}_{k=1}^K}~& \begin{matrix}\sum_{k=1}^K \end{matrix}\text{log}_2\left(1+{(\boldsymbol{\bf{t}}_{k}^T\boldsymbol{\bf{q}}_{k})^2 }\big/{ \sigma^2}\right)\label{o3o}\\ 
\text{s.t.} ~&\|{\boldsymbol{\bf{t}}_{k}}\|^2 \leq P_k, \forall k,\\ 
&\begin{matrix}\sum_{k=1}^K\end{matrix} P_k\leq P,\\
&{\boldsymbol{\bf{t}}_{k}}\succeq  \boldsymbol{0}, \forall k.
\end{split}
\end{equation}

A closer look at \eqref{o3o} shows that for any given feasible  $\{P_k\}_{k=1}^K$, the optimal $\boldsymbol{\bf{t}}_{k}$ can be obtained by applying the Cauchy-Schwarz inequality, given by, $\boldsymbol{\bf{t}}_{k}=\sqrt{P_k}\frac{\boldsymbol{\bf{q}}_{k}}{\|\boldsymbol{\bf{q}}_{k}\|}, \forall k$. As a result, \eqref{o3o} reduces to finding the optimal power allocation ${P_k}$, given by  
\begin{equation}
\begin{split}
\max\limits_{ \{P_k\}_{k=1}^K}~ &\begin{matrix}\sum_{k=1}^K\end{matrix} \text{log}_2\left(1+{P_k\|\boldsymbol{\bf{q}}_{k}\|^2 }\big/{\sigma^2}\right) \\  \label{o4o}
\text{s.t.} ~&\begin{matrix}\sum_{k=1}^K \end{matrix} P_k \leq P,\\
&P_k\geq 0, \forall k.
\end{split}
\end{equation}
It is well known that the optimal solution to \eqref{o4o} is given by the classical WF solution.

\subsection{Path-Based RZF Beamforming} \label{hello}
In this subsection, to achieve a balance between mitigating the interference suffered by MRT beamforming and the noise enhancement suffered by ZF beamforming, we consider the path-based RZF beamforming, where the condition $M_t\geq L_{\mathrm{tot}}$ required by ZF beamforming is no longer needed. Let $\tilde{\boldsymbol{\bf{F}}}=\boldsymbol{\bf{H}}(\boldsymbol{\bf{H}}^H\boldsymbol{\bf{H}}+\epsilon\boldsymbol{I}_{L_{\mathrm{tot}}})^{-1}=[\tilde{\boldsymbol{\bf{f}}}_{11},...,\tilde{\boldsymbol{\bf{f}}}_{1L_1},...,\tilde{\boldsymbol{\bf{f}}}_{K1},...,\tilde{\boldsymbol{\bf{f}}}_{KL_K}]$, where $\epsilon$ is the regularization parameter given by $\epsilon={L_{\mathrm{tot}}\sigma^2}/{P}$ \cite{19}. Then the path-based RZF beamforming is
\begin{equation}
 \boldsymbol{\bf{f}}_{kl}^{\text{RZF}}=\sqrt{p_{kl}}{\tilde{\boldsymbol{\bf{f}}}_{kl}}\big/{\|\tilde{\boldsymbol{\bf{f}}}_{kl}\|}, \label{uu1}
\end{equation}
where $p_{kl}$ is the power allocated to path $l$ of UE $k$.  

With \eqref{uu1}, the desired signal power  in the numerator of  \eqref{45} can be expressed as
\begin{equation}
\begin{split}
P_{\text{DS}}(\boldsymbol{\bf{a}}_k)&=\left|\begin{matrix}\sum_{l=1}^{L_k}\end{matrix}\boldsymbol{\bf{h}}_{kl}^H \boldsymbol{\bf{f}}_{kl}^{\text{RZF}}\right|^2=|\boldsymbol{\bf{a}}_k^T\boldsymbol{\bf{u}}_k|^2\\ \label{pl1}
&=|\boldsymbol{\bf{a}}_k^T\boldsymbol{\bf{u}}_{k,\text{R}}+j\boldsymbol{\bf{a}}_k^T\boldsymbol{\bf{u}}_{k,\text{I}}|^2=\|\boldsymbol{\bf{a}}_k^T\boldsymbol{\bf{U}}_k\|^2,
\end{split}
\end{equation}
where we have defined a non-negative real-valued power allocation vector $\boldsymbol{\bf{a}}_k\triangleq[\sqrt{p_{k1}},...,\sqrt{p_{kL_k}}]^T$, and a complex-valued vector $\boldsymbol{\bf{u}}_k=[\boldsymbol{\bf{h}}_{k1}^H\frac{\tilde{\boldsymbol{\bf{f}}}_{k1}}{\|\tilde{\boldsymbol{\bf{f}}}_{k1}\|},...,$ $\boldsymbol{\bf{h}}_{kL_k}^H\frac{\tilde{\boldsymbol{\bf{f}}}_{kL_k}}{\|\tilde{\boldsymbol{\bf{f}}}_{kL_k}\|}]^H \in \mathbb{C}^{L_k \times 1}$. Furthermore, $\boldsymbol{\bf{u}}_{k,\text{R}}$ and $\boldsymbol{\bf{u}}_{k,\text{I}}$ denote the real and imaginary parts of $\boldsymbol{\bf{u}}_k$, respectively, i.e., $\boldsymbol{\bf{u}}_k=\boldsymbol{\bf{u}}_{k,\text{R}}+j \boldsymbol{\bf{u}}_{k,\text{I}}$, and $\boldsymbol{\bf{U}}_{k}=[\boldsymbol{\bf{u}}_{k,\text{R}}, \boldsymbol{\bf{u}}_{k,\text{I}}]$.
Note that the transformation of \eqref{pl1} to the real-valued vector space facilitates the optimization of the power allocation vector $\boldsymbol{\bf{a}}_k$, which is restricted to the real-valued vector space. Similarly,  the ISI power in the denominator of \eqref{45} can be expressed as
\begin{align}
P_{\text{ISI}}(\boldsymbol{\bf{a}}_k) &=\begin{matrix}\sum_{i=\Delta_{kk,\min},i\ne 0}^{\Delta_{kk,\max}} \end{matrix}\left|\begin{matrix}\sum_{l'= 1}^{L_k}\end{matrix} \boldsymbol{\bf{g}}_{kkl'}^H[i]\boldsymbol{\bf{f}}_{kl'}^{\text{RZF}} \right|^2\\ \notag
 &=\begin{matrix}\sum_{i=\Delta_{kk,\min},i\ne 0}^{\Delta_{kk,\max}}\end{matrix} |\boldsymbol{\bf{a}}_k^T\boldsymbol{\bf{u}}_{kk}[i]|^2\\ \notag
  &=\begin{matrix}\sum_{i=\Delta_{kk,\min},i\ne 0}^{\Delta_{kk,\max}}\end{matrix}|\boldsymbol{\bf{a}}_k^T\boldsymbol{\bf{u}}_{kk,\text{R}}[i]+j\boldsymbol{\bf{a}}_k^T\boldsymbol{\bf{u}}_{kk,\text{I}}[i]|^2\\ \notag
& = \begin{matrix}\sum_{i=\Delta_{kk,\min},i\ne 0}^{\Delta_{kk,\max}} \end{matrix}\|\boldsymbol{\bf{a}}_k^T\boldsymbol{\bf{U}}_{kk}[i]\|^2=\|\boldsymbol{\bf{a}}_k^T\bar{\boldsymbol{\bf{U}}}_{k}\|^2,
\end{align}
where  $\boldsymbol{\bf{u}}_{kk}[i]=[\boldsymbol{\bf{g}}_{kk1}^H[i]\frac{\tilde{\boldsymbol{\bf{f}}}_{k1}}{\|\tilde{\boldsymbol{\bf{f}}}_{k1}\|},..., \boldsymbol{\bf{g}}_{kkL_k}^H[i]\frac{\tilde{\boldsymbol{\bf{f}}}_{kL_k}}{\|\tilde{\boldsymbol{\bf{f}}}_{kL_k}\|}]^H$, $\boldsymbol{\bf{U}}_{kk}[i]$ $= [\boldsymbol{\bf{u}}_{kk,\text{R}}[i], \boldsymbol{\bf{u}}_{kk,\text{I}}[i]]$ and $\bar{\boldsymbol{\bf{U}}}_{k}=[\boldsymbol{\bf{U}}_{kk}[\Delta_{kk,\min}] ,..., $ $ \boldsymbol{\bf{U}}_{kk}[\Delta_{kk,\max}]]$. By following the similar definitions, the IUI power in the denominator of \eqref{45} can be written as $P_{\text{IUI}}(\{\boldsymbol{\bf{a}}_{k'}\}_{k'\neq k}^K)=\sum_{k'\ne k}^{K} \|\boldsymbol{\bf{a}}_{k'}^T\bar{\boldsymbol{\bf{U}}}_{k'}\|^2$. Thus, the SINR \eqref{45} with the path-based RZF beamforming can be expressed as
\begin{equation}
\gamma_k^{\text{RZF}}=\frac{P_{\text{DS}}(\boldsymbol{\bf{a}}_k)}{P_{\text{ISI}}(\boldsymbol{\bf{a}}_k) +P_{\text{IUI}}(\{\boldsymbol{\bf{a}}_{k'}\}_{k'\neq k}^K)+\sigma^2 }. \label{ooo}
\end{equation}

The sum rate can be maximized by optimizing the power allocation vector $\{\boldsymbol{\bf{a}}_k\}_{k=1}^K$ via solving the following problem
\begin{align}
\max\limits_{\{\boldsymbol{\bf{a}}_k\}_{k=1}^K} ~&\sum_{k=1}^K \text{log}_2\left(1+ \frac{P_{\text{DS}}(\boldsymbol{\bf{a}}_k)}{P_{\text{ISI}}(\boldsymbol{\bf{a}}_k) +P_{\text{IUI}}(\{\boldsymbol{\bf{a}}_{k'}\}_{k'\neq k}^K)+\sigma^2 } \right)\label{25}    \nonumber \\ 
\text{ s.t.}~ &\begin{matrix}\sum_{k=1}^K \end{matrix}\|\boldsymbol{\bf{a}}_k\|^2\leq P, \nonumber \\ 
&\boldsymbol{\bf{a}}_k \succeq \boldsymbol{0}, \forall k.
\end{align}
The above problem is non-convex, which cannot be directly solved efficiently. By introducing the slack variables $\bar{\gamma}_k$, problem \eqref{25} can be transformed into
\begin{align}
\max\limits_{\{\boldsymbol{\bf{a}}_k, \bar{\gamma}_k\}_{k=1}^K} ~&\sum_{k=1}^K \text{log}_2(1+ \bar{\gamma}_k  )\label{m1m}  \\ \notag
\text{ s.t.}~&P_{\text{ISI}}(\boldsymbol{\bf{a}}_k) +P_{\text{IUI}}(\{\boldsymbol{\bf{a}}_{k'}\}_{k'\neq k}^K)+\sigma^2\leq \frac{P_{\text{DS}}(\boldsymbol{\bf{a}}_k)}{\bar{\gamma}_k}, \forall k,\\  \notag
& \begin{matrix}\sum_{k=1}^K\end{matrix} \|\boldsymbol{\bf{a}}_k\|^2\leq P, \\ \notag
&\boldsymbol{\bf{a}}_k \succeq \boldsymbol{0}, \forall k.
\end{align}
Though \eqref{m1m} is still non-convex, an efficient Karush-Kuhn-Tucker (KKT) solution can be obtained by using successive convex approximation (SCA) technique.  Specifically, the right-hand-side of the first constraint in \eqref{m1m} is quadratic-over-linear, which is convex and thus is globally lower bounded by its first-order Taylor expansion, i.e., 
\begin{equation}
\frac{P_{\text{DS}}(\boldsymbol{\bf{a}}_k)}{\bar{\gamma}_k}\geq\left(\frac{P_{\text{DS}}(\boldsymbol{\bf{a}}_k)}{\bar{\gamma}_k}\right)_{\text{lb}}\triangleq\frac{1}{\bar{\gamma}_k^r}P_{\text{DS}}(\boldsymbol{\bf{a}}_k^r) + \nabla f(\boldsymbol{\bf{a}}_k^r,\bar{\gamma}_k^r)^T\boldsymbol{\bf{e}}_k,
\end{equation}
where $\boldsymbol{\bf{a}}_k^r$ and $\bar{\gamma}_k^r$ denote the obtained solution at the $r$th iteration, $\nabla f(\boldsymbol{\bf{a}}_k^r,\bar{\gamma}_k^r)=[(\frac{2}{\bar{\gamma}_k^r}\boldsymbol{\bf{U}}_{k}\boldsymbol{\bf{U}}_{k}^T{\boldsymbol{\bf{a}}_k^r})^T,$$ ~\frac{-1}{(\bar{\gamma}_k^r)^2}\|(\boldsymbol{\bf{a}}_k^r)^T\boldsymbol{\bf{U}}_{k}\|^2]^T$ is the gradient, and $\boldsymbol{\bf{e}}_k=[(\boldsymbol{\bf{a}}_k-\boldsymbol{\bf{a}}_k^r)^T, ~\bar{\gamma}_k-\bar{\gamma}_k^r]^T$.
Therefore, for given $\boldsymbol{\bf{a}}_k^r$ and $\bar{\gamma}_k^r$ at the $r$th iteration, the optimal value of \eqref{m1m} is lower-bounded by that of the following problem
\begin{align}
\max\limits_{\{\boldsymbol{\bf{a}}_k, \bar{\gamma}_k\}_{k=1}^K}~& \sum_{k=1}^K \text{log}_2(1+ \bar{\gamma}_k  )\label{m3m}  \\  \notag
\text{ s.t.}~& P_{\text{ISI}}(\boldsymbol{\bf{a}}_k) +P_{\text{IUI}}(\{\boldsymbol{\bf{a}}_{k'}\}_{k'\neq k}^K)+\sigma^2 \leq  \left(\frac{P_{\text{DS}}(\boldsymbol{\bf{a}}_k)}{\bar{\gamma}_k}\right)_{\text{lb}},\\ \notag
& \begin{matrix}\sum_{k=1}^K \end{matrix}\|\boldsymbol{\bf{a}}_k\|^2\leq P, \\  \notag
&\boldsymbol{\bf{a}}_k \succeq 0, \forall k.
\end{align}
Problem \eqref{m3m} is convex, which can be efficiently solved by the standard convex optimization toolbox, such as CVX. By successively update the local point $\{\boldsymbol{\bf{a}}_k^r, \bar{\gamma}_k^r\}_{k=1}^K$, we have the SCA algorithm for \eqref{m1m}, which is summarized in Algorithm \ref{sca}. Note that since the resulting objective value of \eqref{m1m} is non-decreasing over each iteration, Algorithm \ref{sca} is guaranteed to converge.

 \setlength\abovedisplayskip{1.5pt}
\setlength\belowdisplayskip{1.5pt}
 \begin{algorithm}[t]
  \caption{SCA-Based Power Allocation Optimization for Path-Based RZF Beamforming}
  \label{sca}
  \begin{algorithmic}[1]
   \State Initialize a feasible solution $\{\boldsymbol{\bf{a}}_k^0, \bar{\gamma}_k^0\}_{k=1}^K$ to \eqref{m1m}. Let $r=0$;
   \Repeat
   \State {Solve the convex optimization  problem \eqref{m3m} for given }
   \Statex\hspace{5.5mm}{$\{\boldsymbol{\bf{a}}_k^r, \bar{\gamma}_k^r\}$, and denote the optimal solution as $\{\boldsymbol{\bf{a}}_k^{\star}, \bar{\gamma}_k^{\star}\}$. }

   \State \justifying{Update the local point $\boldsymbol{\bf{a}}_k^{r+1}=\boldsymbol{\bf{a}}_k^{\star}$ and $\bar{\gamma}_k^{r+1}=$}$\bar{\gamma}_k^{\star}, \forall k$.
    \State Update $r=r+1$.
    \Until{ The  fractional increase of objective value of \eqref{m3m}  is below a certain threshold}.
  \end{algorithmic}
\end{algorithm}

\section{Simulation Results}

In this section, we provide simulation results to demonstrate the effectiveness of the proposed multi-user DAM communication. We consider a mmWave system at $f_c=$ 28 GHz, with total bandwidth $B=$ 128 MHz, and the  noise power spectrum density $N_0=-174$ dBm/Hz. Thus, the total noise power is $\sigma^2=-93$ dBm. The number of BS antennas and UEs are $M_t=128$ and $K=2$, respectively. Unless otherwise stated, the BS transmit power is $P=30$ dBm, and the number of temporal-resovlable multi-paths for each UE is $L_1=L_2=5$, with their discretized multi-path delays being randomly generated from $[0, 40]$. Besides, the AoDs of all the multi-paths are randomly generated from the interval $[-90^\circ,90^\circ]$, and the complex-valued gains $\alpha_{kl}$ are generated based on the model developed in \cite{3}.
As a benchmarking scheme, the alternative single-carrier scheme  for ISI mitigation  is considered, which only uses the strongest path of each user, termed as the {\it strongest-path} scheme. In this case, the transmitted signal by the BS is  $\boldsymbol{\bf{x}}[n]=\sum_{k=1}^{K}\boldsymbol{\bf{f}}_k s_k[n]$, for which the counterpart MRT, ZF and RZF beamforming can be similarly obtained, by only treating the strongest channel path as the desired path component.

\begin{figure}[t]
\centering
	\includegraphics[scale=0.04]{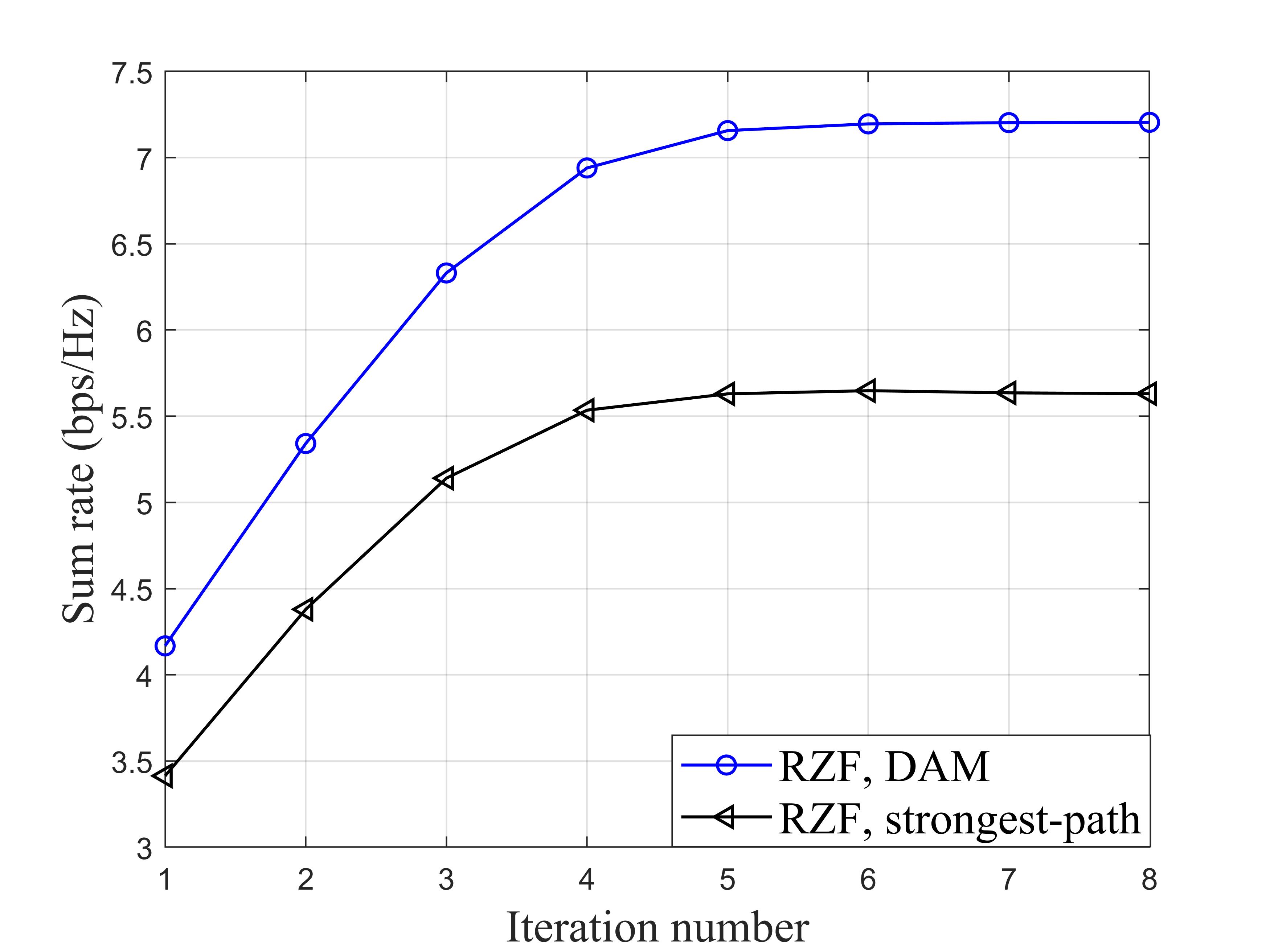}
\caption{Convergence plot of Algorithm 1.}\label{fig3}
 \vspace{-3ex}
\end{figure}

Fig. \ref{fig3} gives the  convergence plot of the SCA-based power allocation optimization for path-based DAM RZF beamforming in Algorithm \ref{sca}, together with the counterpart algorithm for the strongest-path scheme. It is observed from Fig. \ref{fig3} that for both cases, Algorithm \ref{sca} converges quickly with a few iterations.

\begin{figure}[t]
\centering
	\includegraphics[scale=0.04]{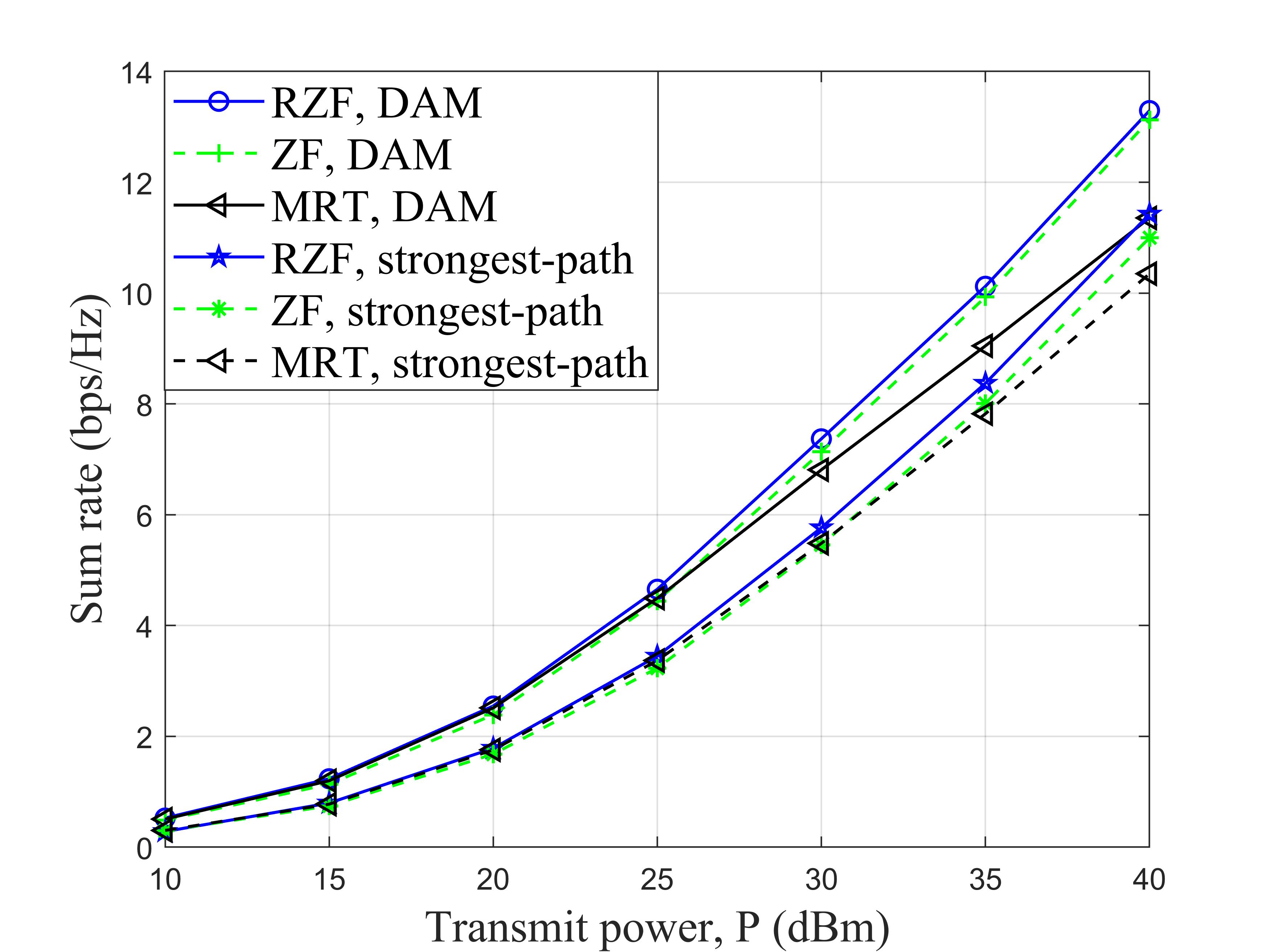}
\caption{Sum rate versus transmit power for the proposed multi-user DAM and the benchmarking  strongest-path scheme.}\label{fig4}
 \vspace{-3ex}
\end{figure}

Fig. \ref{fig4} shows the sum rate versus transmit power for the proposed  multi-user DAM  transmission and the benchmarking strongest-path scheme, with the corresponding MRT, ZF and RZF beamforming, respectively. It is observed from Fig. \ref{fig4} that the proposed multi-user DAM transmission significantly outperforms the benchmarking strongest-path scheme for all the three beamforming schemes. This is expected since DAM makes use of all the multi-path signal components, as can be seen from the first term in \eqref{55}, whereas the strongest-path scheme only uses the strongest multi-path channel component as the desired signal. It is also observed that except for the very  high-power regime ($P\geq 30$ dBm), the low-complexity per-path based MRT beamforming achieves comparable performance as ZF and RZF schemes, thanks to the superior spatial resolution and multi-path sparsity of mmWave massive MIMO systems.

\begin{figure}[t]
\centering
	\includegraphics[scale=0.04]{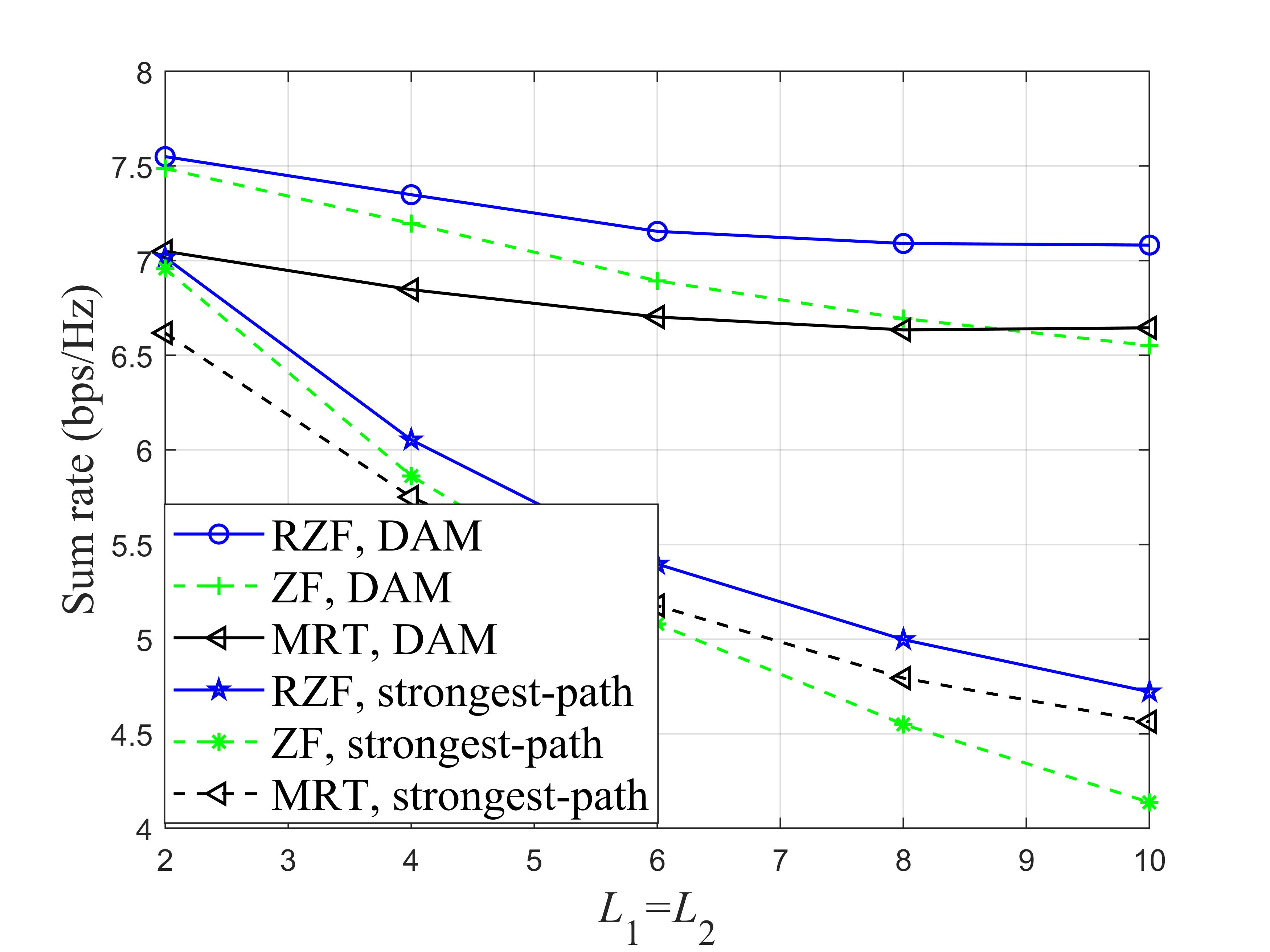}
\caption{Sum rate versus the number of multi-paths for the proposed multi-user DAM and the benchmarking strongest-path  scheme.}\label{fig5}
 \vspace{-3ex}
\end{figure}

Fig. \ref{fig5} studies the impact of the number of  multi-paths on the sum rate for multi-user DAM transmission and the strongest-path scheme. It is observed that the proposed multi-user  DAM transmission shows robustness to the increase of the multi-paths, while the performance of the benchmarking strongest-path scheme degrades significantly with the number of multi-paths. Again, this is because DAM benefits from the delay alignment and constructive superposition  by all the multi-paths, while the strongest-path scheme experiences more severe ISI and IUI as the number of multi-path increases.

\section{Conclusion}
This paper studied the multi-user DAM technique for  mmWave MIMO communication. For the asymptotic case when the number of BS antennas is much larger than the total number of channel multi-paths, we showed that the ISI and IUI can be completely eliminated with the simple path-based MRT beamforming and delay pre-compensation. Then for the more general multi-user DAM design, three classical beamforming schemes in a per-path basis tailored for DAM communication were investigated, namely the path-based MRT, ZF and RZF beamforming. Simulation results demonstrated that the proposed multi-user DAM  outperforms the benchmarking single-carrier ISI mitigation technique that only uses the strongest channel path of each user.

\section*{Acknowledgment}
This work was supported by the National Key R\&D Program of China with Grant number 2019YFB1803400.


\end{document}